\documentclass[%
reprint,
11pt,
nofootinbib,
amsmath,amssymb,
aps,
]{revtex4-1}

\usepackage[utf8]{inputenc}
\usepackage[T1]{fontenc}
\usepackage{fontsize}
\changefontsize[11]{10}
\usepackage{graphicx}
\usepackage{dcolumn}
\usepackage{bm}
\usepackage{ragged2e}
\usepackage{amsmath,amssymb} 
\usepackage{mathrsfs}
\usepackage{siunitx}
\usepackage{arydshln}
\usepackage{amsfonts}
\usepackage{pgfplots}
\usepackage{floatrow}
\usepackage{float}
\usepackage{caption}
\usepackage{subfig}
\usepackage{xcolor}
\usepackage[symbol]{footmisc} 
\usepackage[utf8]{inputenc}
\setcitestyle{super}     

\makeatletter
\renewcommand\citet[1]{\citeauthor{#1}\textsuperscript{\cite{#1}}}
\makeatother

\definecolor{Green}{HTML}{167425}
\definecolor{azure}{HTML}{003FFF}

\usepackage[colorlinks=true,linkcolor=black,citecolor=black]{hyperref}%

\usepackage[normalem]{ulem}

\def\IB#1{\boldsymbol{#1}} 
\def\W/!i#1{\Wi} 

\newcommand{\rev}[1]{\textcolor{black}{#1}}

\begin{document}
\title{
On the influence of electrolytic gradient orientation on phoretic transport in dead-end pores
}
\author{Kushagra Tiwari}\thanks{Equal contribution}
\author{Jitendra Dhakar}\thanks{Equal contribution}
\author{Kapil Upadhyaya}
\author{Akash Choudhary}
\email{\textcolor{black}{Corresponding author:} \textcolor{black}{achoudhary@iitk.ac.in}}
\address{Department of Chemical Engineering, Indian Institute of Technology Kanpur, 208016, India}

    \begin{abstract}
\noindent
\textbf{Abstract.} Electrolytic diffusiophoresis refers to directional migration of colloids due to interfacial forces that develop in response to local electrolytic concentration ($c$) gradients.
This physicochemical transport provides an efficient alternative in numerous microscale applications where advection-induced transport is infeasible.
	Phoretic withdrawal and injection in dead-end pores can be controlled by orienting salt gradients into or out of the pore; however, the extent to which this orientation influences spatiotemporal transport patterns is not thoroughly explored. 
    In this study, we find that it has a significant influence: colloidal withdrawal in solute-out mode ($\beta=c_\infty/c_{\text{pore}}<1$) is faster and shallower, whereas the solute-in mode enables deeper withdrawal.
    Similarly, solute-out injection features rapidly propagating wavefronts, whereas the solute-in mode ($\beta >1$) promotes uniform and gradual injection. 
Each mode's transport is found to evolve and persist over different time scales. 
We characterize the performance of these modes and find that while persistence of the solute-out mode strengthens with a growing electrolytic gradient [$\sim \ln(\beta^{-0.4})$], solute-in mode diminishes and eventually its persistence is insensitive to $\beta$.
We also incorporate the variable mobility model to examine the impact of large zeta potentials, which intensifies the transport of solute-out mode further and weakens the solute-in mode. 
Additionally, we investigate how osmotic flows of the two modes affect injection and withdrawal patterns. We find that osmosis-induced mixing can counterintuitively inhibit injection effectiveness in solute-out mode. These insights bring attention to the distinctions between different phoretic transport modes and contribute to the rational design and setup of electrolytic gradients in numerous microscale applications.
    \end{abstract}
    \maketitle

\section*{Introduction}

 	\begin{table*}[t]
  \small
     		\renewcommand{\arraystretch}{0.85}
 		\begin{center}
 		\def~{\hphantom{0}}
\begin{tabular}{l l l l}
\textbf{Investigation} & $\qquad$\textbf{Description} & \textbf{Conditions/Regime} & \textbf{Key Results/Insights} \\[8pt] \noalign{\hrule height 0.4pt}
\\
			\begin{tabular}[t]{l}  Kar \textit{et al.} $\;\;\; \;$\\(2015) \cite{kar2015enhanced} \end{tabular}       & \begin{tabular}[t]{l}Exp. studied colloids \\transport into and out\\of dead-end pore\end{tabular}      & \begin{tabular}[t]{c}  3 - 4 $\mu$m sized polystyrene\\(PS) beads and emulsions \end{tabular}     & \begin{tabular}[t]{l} Transient DP and DO flow governs\\colloidal transport in dead-end pores\end{tabular} \\ \\
   
            \begin{tabular}[t]{l} Shin \textit{et al.}\\(2016) \cite{shin2016size} \end{tabular}       & \begin{tabular}[t]{l}Exp. and num. studied \\size-dependent control of\\colloidal transport  \end{tabular}      & \begin{tabular}[t]{l} $0.06-1.01$ $\mu$m sized PS,\\latex particles; $\zeta_p$ constant\end{tabular}     & \begin{tabular}[t]{l} Devised the first continuous exp. setup.\\Larger particles inject deeper due to\\higher mobility \end{tabular} \\ \\
            
            \begin{tabular}[t]{l} Ault \textit{et al.}\\(2017) \cite{ault2017diffusiophoresis} \end{tabular}       & \begin{tabular}[t]{l} Analytically and num.\\studied 1D injection\\\& withdrawal transport\end{tabular}      & \begin{tabular}[t]{l} $\zeta_p =$ constant, $\text{Pe} \to 0$ 
            \\ Studied two regimes \\ \textit{(i.)}  $x \ll L $, \\ \textit{(ii.)} $x \sim L $ for $\beta \to 0$ \end{tabular}     & \begin{tabular}[t]{l} Analytical results in diffusion-free limit.\\Results for \textit{(i)} semi-infinite \&\textit{(ii)} finite\\domains. Insights into colloid injection\\dynamics. \end{tabular} \\ \\

            \begin{tabular}[t]{l} Shin \textit{et al.}\\(2017) \cite{shin2017low} \end{tabular}       & \begin{tabular}[t]{l} Exp. devised a low-cost\\potentiometry device  \end{tabular}      & \begin{tabular}[t]{l} Functionalized polystyrene \\particles and vesicles\end{tabular}     & \begin{tabular}[t]{l} Technique to measure zeta potentials\\by tracking phoretic \& osmotic transport \end{tabular} \\ \\
            \begin{tabular}[t]{l} Wilson \textit{et al.}\\(2020) \cite{wilson2020diffusiophoresis}  \end{tabular}       & \begin{tabular}[t]{l} Exp. studied DP in multi\\valent electrolyte gradient\end{tabular}      & \begin{tabular}[t]{l} 1 $\mu$m polystyrene particles\\in 6 sets of electrolytes \end{tabular}     & \begin{tabular}[t]{l} Injection and compaction studies showed\\the impact of cation/anion diffusivity.\end{tabular} \\ \\
            
            \begin{tabular}[t]{l} Gupta \textit{et al.}\\(2020) \cite{gupta2020diffusiophoresis}\end{tabular}       & \begin{tabular}[t]{l} Num. and exp. studied\\the effects of electrolyte\\concentration on DP \end{tabular}      & \begin{tabular}[t]{l}  500 nm polystyrene part-\\icle with $\zeta_p=f(c)$,\\$c^\ast \rightarrow 0.1-1000$, $\beta=0.1$\end{tabular}     & \begin{tabular}[t]{l} Reported that optimum DP transport\\ occurs between $c^\ast \sim$ $\mathcal{O}(1)$ to $\mathcal{O}(10)$ mM \end{tabular} \\ \\

            \begin{tabular}[t]{l} Alessio \textit{et al.}\\({2022}) \cite{alessio2022diffusioosmosis}\end{tabular}       & \begin{tabular}[t]{l} Exp. and num. showed\\3D effects on colloidal\\dispersion \end{tabular}      & \begin{tabular}[t]{l} 0.2-1 $\mu$m polystyrene and\\carboxylate particles, $\text{Pe} \lesssim 1$ \end{tabular}     & \begin{tabular}[t]{l} DO from sidewalls governs dispersion\\\& formulated effective dispersion coe-\\fficient \end{tabular} \\ \\

            \begin{tabular}[t]{l} Akdeniz \textit{et}\\\textit{al.}(2023)\cite{akdeniz2023diffusiophoresis}\end{tabular}       & \begin{tabular}[t]{l}Exp. and num. examined\\the effects of concentration- $\;\;$\\dependent zeta potential \end{tabular}  & \begin{tabular}[t]{l} $\sim$ 1 $\mu$m sized PS-PEG \&\\carboxylate particles,\\$\zeta_p(c)$ \& $\zeta_w(c)$, $\text{Pe}\lesssim 1$\end{tabular}     & \begin{tabular}[t]{l} Variable-mobility model calculations\\agree well with exp. over long time scales.\\
            Impact of advective flow at pore entry.\end{tabular} \\ \\
            
            \begin{tabular}[t]{l} Lee \textit{et al.}\\(2023) \cite{lee2023role}\end{tabular}       & \begin{tabular}[t]{l} Num. studied the impact\\of variable mobility on\\DP and DO transport \end{tabular}      & \begin{tabular}[t]{l} Reassessed  refs.\cite{staffeld1989diffusion,ault2017diffusiophoresis,ault2019characterization,chu2020advective}\\ for $ \zeta_p=f(c)$ and $\text{Pe} \to 0$  
            \end{tabular}     & \begin{tabular}[t]{l} Found that reference zeta potential for\\constant mobility model can be derived\\using the constant charge model.  \end{tabular} \\ \\

            \begin{tabular}[t]{l} Migacz \textit{et al.}\\(2024) \cite{migacz2024enhanced} \end{tabular}       & \begin{tabular}[t]{l}Studied the effects of time-\\dependent ambient solute\\boundary condition \end{tabular}      & \begin{tabular}[t]{l}  $\text{Pe} \to 0$, $\beta=f(t)$: linear\\and oscillatory\end{tabular}     & \begin{tabular}[t]{l} Slow variations yield linear injection/\\withdrawal, fast \& optimal variations\\yield non-linear \& enhanced transport   \end{tabular} \\ \\
            
            \begin{tabular}[t]{c} This work \end{tabular}       
& \begin{tabular}[t]{l} 
   Num. studied the impact of\\
   solute gradient orientation\\
   on withdrawal \& injection\\
   performance 
   \end{tabular}      
& \begin{tabular}[t]{l} 
   $\;\;$(\textit{i})  1D: $\zeta_p$ constant, $\text{Pe} \to 0 \;\;$\\
   $\;\;$(\textit{ii}) 1D: $\zeta_p(c), \text{Pe} \to 0 \;\;$\\
   $\;\;$(\textit{iii}) 2D: $\zeta_p(c)$, $\zeta_w(c)$, $\text{Pe}\lesssim 1 \;\;$
   \end{tabular}             
& \begin{tabular}[t]{l} 
   Solute-out \& -in modes exhibit disparate\\
   (\textit{i}) spatiotemporal evolution \&  (\textit{ii})\\
  scaling of performance w.r.t gradient $(\beta)$.\\
  DO amplifies effectiveness, except solute-\\out injection
   \end{tabular}
      \\ \noalign{\hrule height 0.4pt}
            \end{tabular}
        \caption{Past studies on transport of colloids into and out of the dead-end pore driven by diffusiophoresis and diffusioosmosis. Here, $\zeta_p$ (mV), $\zeta_w$ (mV), $c^\ast$ (mM), and $\beta$ are the particle zeta potential, wall zeta potential, solute concentration in the pore, and solute concentration ratio between reservoir and pore, respectively. (Exp. - experimentally, Num. - numerically).
        Here Pe is the solute Peclet number, DP and DO represent diffusiophoresis and diffusioomsosis, respectively. \rev{Here $L$ denotes the pore length.}} 
 		\label{table1}
 		\end{center}
 	\end{table*}

    \begin{figure*}[t]
    \centering
    {\includegraphics[width=1\linewidth]{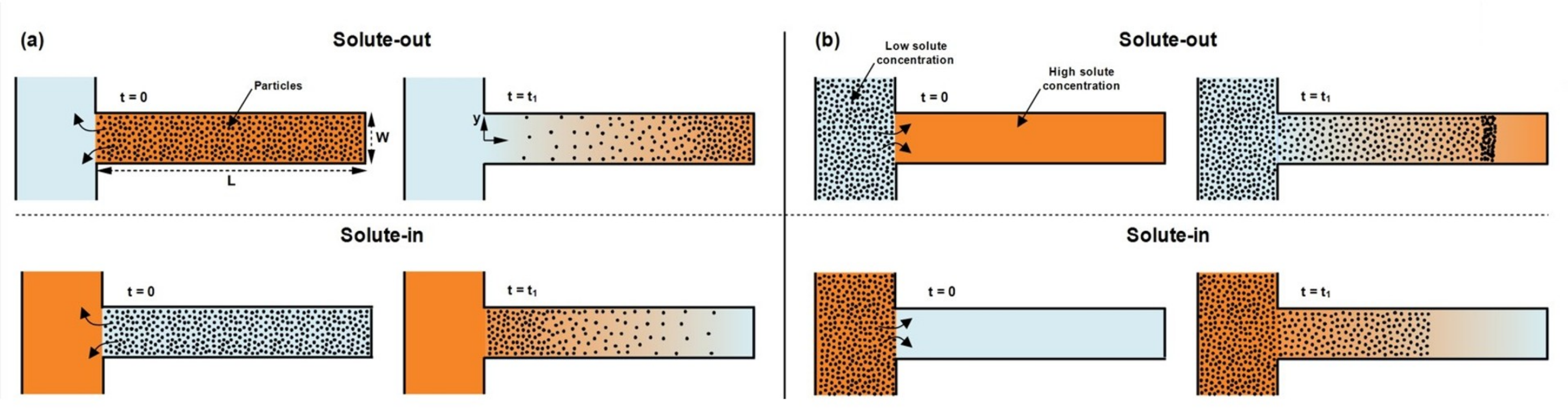}}
    \caption{Schematic of particle (a) withdrawal and (b) injection dynamics by different solute gradient orientations. 
    Here, $\beta$ is a ratio representing relative concentrations between reservoir and pore.
   }
    \label{fig:schematic}
    \end{figure*}

\noindent
Microscopic colloids in a molecular bath experience potential forces from intermolecular interactions within the Debye screening region near their surface. External concentration gradients disrupt the equilibrium of thermal and interactive forces, inducing a diffusioosmotic (DO) slip velocity. This slip on a freely suspended colloid drives a diffusiophoretic (DP) motion, where it migrates parallel to chemical gradients $\propto \nabla c$ \cite{derjaguin1947kinetic,anderson1989}.
When the solute is an electrolyte, differential ion diffusion creates electrostatic gradients ($\nabla \psi$) alongside chemical gradients. This generates a local electric field that modifies the osmotic/phoretic contribution, proportional to $\nabla \log c$ \cite{prieve1984motion,velegol2016origins,ault2024physicochemical}. 
This motion is mediated by diffusive and electrostatic solute-surface interactions, encapsulated in chemo- and electro-phoretic mobility parameters.

The resurgence of these phoretic and osmotic phenomena after decades of initial work is attributed to the rise of micro- and nano-fluidic technologies, which enable precise control over fluid and particle transport \cite{shi2016diffusiophoretic}. 
\rev{
Through experiments on focusing and spreading of colloids, the seminal study by \citet{abecassis2008} demonstrated two orders-of-magnitude enhancement in effective `diffusion' in salt gradients $\sim \Gamma_p^2/D_s$ ($\Gamma_p$ and $D_s$ are phoretic mobility and salt diffusion coefficient). 
Later, \citet{palacci2010colloidal} built microfluidic devices for patterned trapping of colloids and $\lambda-$DNA in temporally-controlled soute concentration profile.
In response to salt oscillations at varying frequencies, the particles settled into a characteristic Gaussian distribution. The width of this distribution depended on the phoretic mobility, which varied with different salts (LiCl > NaCl > KCl), as it is influenced by the difference in the diffusivities of the dissociated ions.
Later, \citet{palacci2012osmotic} analysed the time scales and robust persistence of colloidal migration via experiments and numerical simulations. 
They demonstrated that phoretic log-sensing enables efficient particle focusing across various ``osmotic shock'' configurations within a microfluidic circular well chamber.
}
These physicochemical effects can also be leveraged for designing bioassays that trap vesicles  \cite{rasmussen2020size} \& transport DNA \cite{katzmeier2024reversible}, performing low-cost zeta potentiometry \cite{shin2017low,ault2019characterization}, and designing hydrophobic nanopores that can act as osmotic diodes for rectified transport \cite{picallo2013nanofluidic}.
Furthermore, potential \textit{discoveries} of osmotic and phoretic phenomena have recently emerged within applications and technologies that have existed for decades. 
Examples include, dirt transport in constricted fibre pores during conventional laundering \cite{shin2018cleaning},  phoresis-induced fouling, osmotic flows in  water filtration membranes \cite{sablani2001concentration,shin2017accumulation}, and  low-salinity water flooding for tertiary oil recovery \cite{kar2015enhanced, velegol2016origins,lager2008losal}. 
\rev{\citet{lee2018diffusiophoretic} fabricated a microfluidic device for continuous water purification in conduits made from ion-exchange membranes and studied exclusion of particle and pathogens in a developing ionic boundary layer.
\citet{bekir2023versatile} employed light-induced diffusioosmosis on particle surfaces to yield a lift, which aided in separating particles on the basis of size and porosity.
\citet{chakra2023continuous} devised a continuous experimental procedure to sort and characterize nanoparticles and liposomes on a $\Psi$-junction microfluidic chip.
 }

Most of these applications involve transport of colloids within stagnant dead-end pores that facilitate negligible fluid convection. Hence, electrolytic DP can serve as an efficient method for microporous transport. To set this up as colloidal injection or withdrawal, the chemical gradients can be applied in either `solute-out' mode (solute-saturated pore meets solute-free ambient) or `solute-in' mode (solute-free pore meets solute-saturated ambient).
In their seminal experimental study, \citet{kar2015enhanced} and \citet{shin2016size} demonstrated how droplets and particles can be efficiently transported in dead-end pores in a controlled manner. 
Using a continuum framework, \citet{ault2017diffusiophoresis} later focused on modeling the colloidal  injection and withdrawal dynamics and showed that colloids enter the pores as propagating bands. In the limit of times smaller than characteristic solute diffusion, zero colloidal diffusivity, and weak phoretic mobility, they found that injection wavefronts develop as $\sim t^{1/2}$ for solute-out mode, whereas solute-in mode dynamics show weak temporal development.
Later, \citet{gupta2020diffusiophoresis} showed the impact of solute concentration on mobility and solute-out compaction, emphasizing the importance of finite-Debye layer effects for transport of sub-micron sized colloids.
As part of their study, \citet{lee2023role} further extended the above two studies (for solute-out injection) to large zeta potentials, such that the colloid's phoretic mobility varies with local electrolyte concentration, i.e., the constant charge (or variable mobility) model. 
Of the two effects, finite-Debye layer effects have a greater impact on the quantitative nature of the results, whereas their influence on the qualitative nature of the injection profile is negligible.
\rev{
Recently, \citet{tan2021two} demonstrated a two-step process for persistent delivery of colloids in porous media by utilizing targeted dead-end pores—i.e., pore ends laden with high-solute-absorption-capacity, such as Polyethylene Glycol Diacrylate (PEG-DA).
In the first step, the entire pore (including the target) are saturated with solute. 
In the second step, a solute-free colloidal suspension encounters the pore mouth. Due to the high partition coefficient of PEG-DA target, it acts as a beacon, slowly releasing solute and maintaining long-lasting gradients that facilitate deeper colloid injection.
}

In addition to particle diffusiophoresis, bulk osmotic flow can shape the particle dispersion because channel or pore walls are generally charged and, in the presence of external solute gradients, can induce diffusio-osmotic slip velocity \cite{kar2015enhanced,shin2016size}.
This slip velocity drives a nonuniform flow within the pores to maintain zero net flow. 
Using Taylor dispersion analysis and experiments, \citet{alessio2022diffusioosmosis} further demonstrated that osmotic flows originating from side walls of the channel can have a significant impact on colloidal dispersion.
Recently, a similar experimental and numerical study by \citet{akdeniz2023diffusiophoresis} highlighted the impact of implementing constant-charge model on colloidal dispersion, particularly over long timescales.
Table \ref{table1} provides an overview of these significant contributions towards understanding colloidal transport in dead-end pore.

Since the first experiments and computations \cite{kar2015enhanced,shin2016size,ault2017diffusiophoresis} on dead-end pores, significant progress has been made in understanding the impact of multiple and multivalent electrolyte gradients \cite{alessio2021diffusiophoresis,gupta2019diffusiophoretic}, pH gradients \cite{shim2022diffusiophoresis}, photocatalytic particles \cite{visan2019reaction}, temporal control of solutal conditions \cite{ha2019dynamic,migacz2024enhanced}, and polymeric gradients \cite{akdeniz2024diffusiophoresis}. 
However, these studies have  focused primarily on colloidal injection and compaction in the solute-out mode, as they provide the most pronounced experimental results in short timescales, leaving the other three cases largely unexplored.
Furthermore, phoretic transport in naturally occurring scenarios \cite{lager2008losal, ganguly2023diffusiophoresis, ault2024physicochemical} may not be limited to the solute-out mode.
In this work, we examine previously overlooked qualitative differences and demonstrate that solute-in and solute-out modes of colloidal transport exhibit distinct spatiotemporal dynamics as well as performance characteristics; a preliminary insight can be found in Fig. \ref{fig:schematic}.
Specifically, this study is driven by the following questions:
\begin{enumerate}
    \item How and why does the evolution of colloidal distribution differ for the two orientational modes?
    \item How do these modes compare in terms of injection/withdrawal performance: phoretic lifespan and effectiveness?
    \item How do the osmotic flows of the two modes differ, and how is the performance further altered?
\end{enumerate}


In what follows, we describe the continuum framework used to predict and obtain qualitative insights into colloidal transport. 
\rev{Since the two modes evolve over vastly different timescales, their lifespans cannot be characterized by a common phoretic timescale $L^2/\Gamma_p$. We therefore quantify performance using two separate metrics:} persistence time and effectiveness.
Subsequently, we explore the influence of variable mobility model on spatiotemporal patterns.
Additionally, we investigate how osmotic flows affect injection and withdrawal dispersion, finding that osmosis-induced mixing can counterintuitively inhibit injection effectiveness in solute-out mode.
Finally, we discuss key conclusions and outlook.

\section*{Materials and Methods}
\textbf{Mathematical Model.}
We consider the diffusiophoretic motion of freely suspended and non-interacting colloids in dead-end pores driven by the two orientational modes of solute gradients. 
We consider gradients of  symmetric  electrolytes, where the ionic concentration is represented by $c(x,t)$. 
Fig. \ref{fig:schematic} (a) depicts the case when colloids are withdrawn from a pore, whereas Fig. \ref{fig:schematic} (b) shows the injection.
At the continuum scale, this coupled transport is governed by the following convection-diffusion equations for solutal and colloidal dynamics \cite{shin2016size}
\begin{gather}
		\frac{\partial c}{\partial t} +  \IB{\nabla}\cdot \left[\text{Pe}\, \IB{u}_\text{f} c\right] =  {\IB{\nabla}}^2 c ,\label{c-GE}
        \\
		\frac{\partial n}{\partial t} + \IB{\nabla} \cdot \left[\text{Pe} \, \IB{u}_\text{f} \, n  +  \IB{u}_\text{p} \, n  \right] =  \mathcal{D}  {\IB{\nabla}}^2 n .
        \label{n-GE}
	\end{gather}  
These equations have been non-dimensionalized using $c_\text{i}$, $n_\text{i}$, $L$, $L^2/D_\text{s}$, {$\Gamma_w/L$}, and {$\mu \Gamma_w/L^2$} for solute concentration, particle concentration, \rev{pore} length, time, velocity, and pressure, respectively.
The normalized phoretic velocity $\IB{u}_p = \frac{\Gamma_p}{D_\text{s}} \IB{\nabla} \ln c$ couples the two equations.
These equations are complemented with the quasi-steady flow Stokes equations that govern the 2D osmotic flow emerging from pore walls: $\IB{\nabla}\cdot \IB{u_\text{f}} = 0 $ and $0 = \IB{\nabla} p +  {\IB{\nabla}}^2 \IB{u_\text{f}}$.
The solute Peclet number {$\text{Pe} = \Gamma_w/D_s$} captures the ratio of osmotic flow convection relative to solute diffusion.  We have defined the dimensionless $\mathcal{D}$ for capturing colloidal diffusion ($D_\text{p}/D_\text{s}$) relative to the solute diffusion.
Here, $c_\text{i}$ represents the initial solute concentration in the pore (reservoir) for solute-out (solute-in) mode, and similarly, $n_\text{i}$ represents the initial colloidal concentration in the pore (reservoir) for withdrawal (injection).

In dimensionless form, the initial conditions for the solute are given by  $c(x, t=0) = 1$  for the solute-out mode and  $c(x, t=0) = 0$  for the solute-in mode. The boundary conditions are specified as  $c(x=0, t) = \beta$  at the inlet and a no-flux condition at  $x = 1$. Here, $\beta$ denotes the relative concentration ratio between the reservoir and the pore, with $\beta < 1$ for solute-out and $\beta > 1 $ for solute-in.
Similarly, the conditions for colloids are $ [n(x,t=0)= 1 \; \& \; n(x=0,t) = 0] $ and $ [n(x,t=0)= 0 \; \& \; n(x=0,t) = 1] $ for withdrawal and injection transport, respectively (with no-flux condition at $x=1$). 
\rev{Finally, the velocity boundary condition utilized at the pore walls is the diffusioosmotic slip$-\frac{\Gamma_w}{D_\text{s}} {\nabla} \ln c$, and following earlier studies \cite{shin2016size,akdeniz2023diffusiophoresis}, zero pressure \& zero net volume flux conditions 
are used.}
These conditions implicitly assume that there are no solutal/colloidal transport resistances in the reservoir, i.e., the ambient flow is fast enough to maintain these conditions at all times.

        \begin{figure*}[t]
		\centering
        {\includegraphics[width=1.05\linewidth]{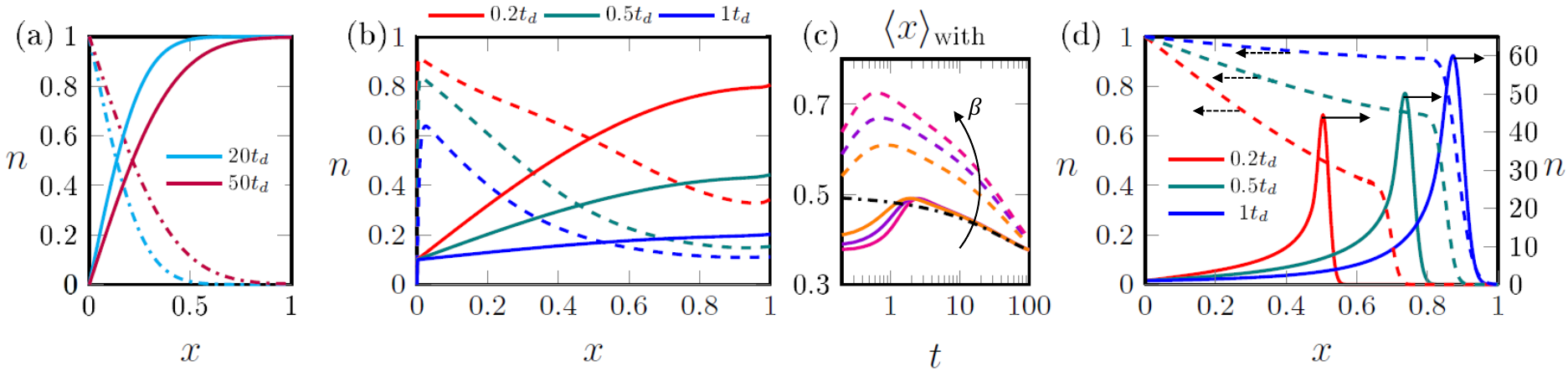}}
		\caption{\textbf{Spatio-temporal profile of colloids.}
        (a) Pure diffusion profile ($\Gamma_p=0$) for injection (\rev{dash-dotted}) and withdrawal (\rev{solid}).
        (b) Dynamics of colloid withdrawal, where solid lines depict shallow withdrawal in solute-out mode and dashed lines depict deeper withdrawal in solute-in mode. (c) Weighted average position of withdrawn population for $\beta=0.05,0.1,0.2,5,10,20$. \rev{Here, solid and dashed lines depict the solute-out and solute-in modes, respectively.}
        The dash-dotted line represents bare colloid diffusion for $\beta=1$. 
        (d) Injection dynamics. The arrows depict the axis each mode corresponds to.  Parameters: $D_s=10^{-9}$ m$^2$/s, $D_p=10^{-12}$ m$^2$/s, $\Gamma_p=10^{-9}$ m$^2$/s, $\beta=0.1$ (solute-out, solid lines), and $\beta=10$ (solute-in, dashed lines). }
		\label{fig:with_inj}
	\end{figure*}

    \textbf{Computational Method.} Solute concentration can be derived analytically using separation of variables \cite{shin2016size,ault2017diffusiophoresis}
\begin{equation}
    c(x,t) = \beta + (1 - \beta) \sum_{n=0}^{\infty} \frac{2(1 - \cos \lambda_n)}{\lambda_n} \sin\left({\lambda_n x}\right) e^{-\lambda_n^2 t},
\end{equation}
\rev{where $\lambda_n= \pi (2n+1)/2$.}
We use the first 10\textsuperscript{3} terms of this series to evaluate the phoretic velocity.
\rev{We note that the step change in solute concentration at pore entry corresponds to a singularity in phoretic velocity at $t=0$. 
For practical purposes, we focus on the times $t^*\geq 0.1 t_d$ in all our numerical results. In employing the series solution for solute, we use the first 10\textsuperscript{3} terms to ensure excellent accuracy.}
 The 1D colloidal concentration profile was obtained numerically by employing the Method of lines, that utilized second order finite-difference discretization in space \rev{($\Delta x=0.001$ \& $\Delta t=0.005$)}. 
 In addition to the grid independence test, we verified our results using COMSOL multiphysics (see \S 1 in Supplementary material), which is also used to obtain the 2D profile. Therein, we also compare our results with earlier numerical studies \cite{shin2016size,lee2023role}.

	\section*{Results and Discussion}

 We begin the discussion with the assumption that colloids and pore walls exhibit uniform and constant mobilities. Although zeta potentials can be sensitive to local concentrations \cite{kirby2004zeta}, we first focus on building a systemic understanding and relax this assumption in a later subsection.
Furthermore, to isolate the phoretic effects for clarity, we consider zero mobility of the confining pore walls, i.e., absence of diffusioosmotic flow ($\text{Pe} \to 0$), which renders the system primarily 1D transient. 
	This assumption will also be relaxed in the subsequent subsections.
\rev{In the results below, we consider colloids with both positive and negative phoretic mobilities, which can arise for either positive or negative zeta potentials, provided the ion diffusivity difference, defined as \(\beta_s = \frac{D_+ - D_-}{D_+ + D_-}\), has an appropriate sign and magnitude. Using the mobility expression from \citet{prieve1984motion} and assuming room temperature properties of water, estimates from the experimental studies of \citet{shin2017low} suggest that a micron-sized native polystyrene particle (\(\zeta_p = -79\, \text{mV}\)) yields a mobility of \(\Gamma_p = 0.95 \times 10^{-9}\, \text{m}^2/\text{s}\) in NaCl (\(\beta_s = -0.208\)), whereas in SDS (\(\beta_s = 0.55\)), the mobility is approximately \(-0.26 \times 10^{-9}\, \text{m}^2/\text{s}\).
Although positive mobility is more common in neutral electrolytes like NaCl, polystyrene particles in concentrated HCl mediums ($\zeta_p = -34$ mV, $\beta_s=0.642$) can yield a negative mobility of $-0.33 * 10^{-9}$ m\textsuperscript{2}/s
\cite{shim2022diffusiophoresis}.
Furthermore, amine-modified polystyrene particles can exhibit positive zeta potentials ($\zeta_p \approx +43$ mV) that correspond to negative mobilities of $-0.34 * 10^{-9}$ m\textsuperscript{2}/s in NaOH gradients ($\beta_s=-0.596$) \cite{shim2022diffusiophoresis}.
Following a similar rationale, we also consider both positive and negative mobilities for pore walls in a later subsection. Although direct experimental evidence for positive wall zeta potentials or negative wall mobilities is currently lacking, such conditions are theoretically plausible based on the particle-level arguments. 
We note that cases involving particles and walls with opposite zeta potentials are experimentally challenging due to electrostatic-interaction-induced fouling~\cite{breite2016critical}.}

\subsection*{Phoretic withdrawal and injection for constant mobility}

\textit{Withdrawal.} First, we explore the withdrawal of colloids that are trapped within the pore at time $t=0$, as depicted in Fig.\ref{fig:schematic}(a). Fig.\ref{fig:with_inj}(a) shows the spatio-temporal colloid density profiles for pure colloidal diffusion ($\Gamma_p=0$). It requires $10^2$ orders of solute diffusion time ($t_d$) to withdraw a substantial amount of particles.
 Fig.\ \ref{fig:with_inj}(b) shows that diffusiophoresis can offer similar magnitudes of withdrawal in $\sim t_d$ times. Estimations of such an intensified transport in dead-end pores have been explored earlier \cite{palacci2010colloidal,palacci2012osmotic,lee2023role,gupta2020diffusiophoresis,ault2024physicochemical}. 
Here, we focus on the influence electrolytic-gradient direction has on the qualitative and quantitative aspects.
The trends in Fig.\ref{fig:with_inj}(b) depict that solute-out mode offers a shallow withdrawal where colloids from pore-end experience delayed entrainment, whereas solute-in mode facilitates a deeper withdrawal. We quantify these trends in Fig.\ \ref{fig:with_inj}(c) which shows the weighted average position of the withdrawn colloids: $\langle x \rangle_{\text{with}} = 1 - \left({\int_{0}^{1} n \, x \, \text{d}x}/ {\int_{0}^{1} n \, \text{d}x} \right)$, where the measure for solute-in (-out) mode lies above (below) $x = 0.5$. Post diffusiophoretic lifespan (after few $t_d$), both trends asymptotically approach the diffusion-mediated colloid withdrawal shown by the dash-dotted curve for $\beta =1$.

    \begin{figure}[t]
		\centering
         {\includegraphics[width=1.03\linewidth]{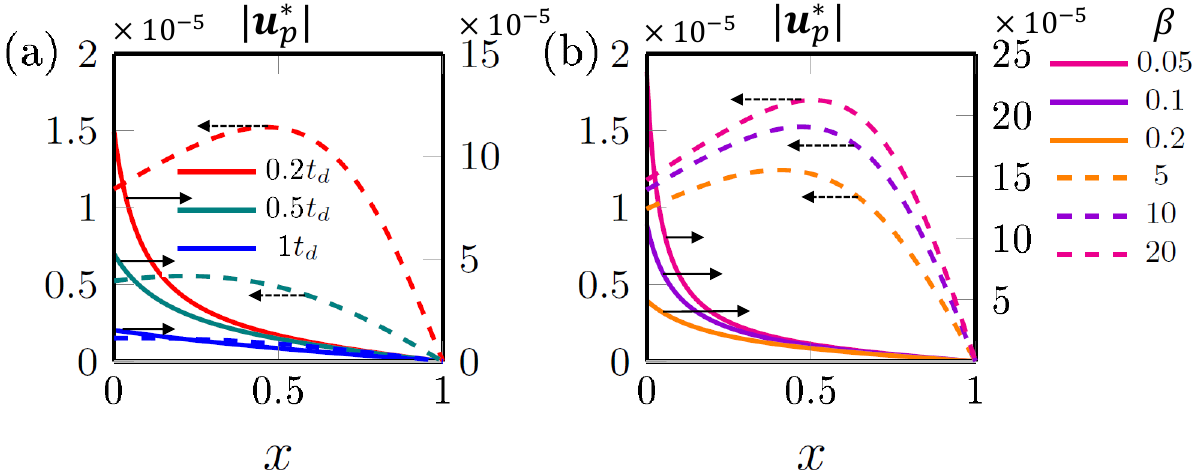}}
	\caption{\textbf{Spatial variation of phoretic velocity.}
    (a) Solute-out  ($\beta=0.1$, solid lines) and solute-in  ($\beta=10$, dashed lines) modes at different times.  The arrows point to the axis to which each mode corresponds. (b) Variation for different solute gradients $\beta$ at time 0.2$t_d$. Parameters: {$D_s=10^{-9}$ $m^2$/s and $\Gamma_p=10^{-9}$ m$^2$/s.}}
	\label{fig:Vel_cz}
	\end{figure}

        \begin{figure*}[t]
		\centering
        {\includegraphics[width=1\linewidth]{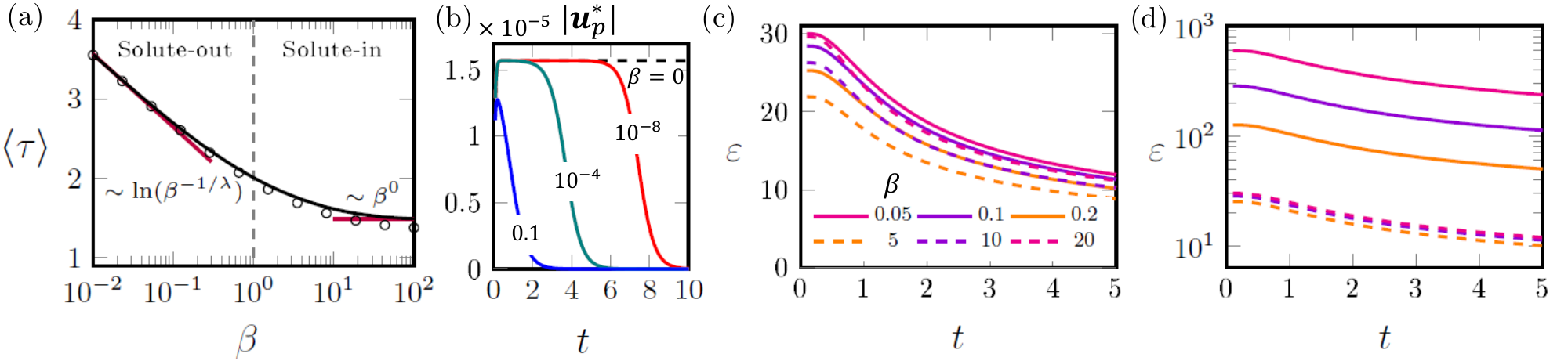}}
\caption{\textbf{Performance measures.} (a) Variation of phoretic persistence time with applied concentration ratio $\beta$. The \rev{purple} lines \rev{at the extremes} show the analytical predictions from Eq. (\ref{analytical}); $\lambda=\pi^2/4$. Persistence is identical for colloid injection and withdrawal. \rev{The empty circles correspond to Eq. (\ref{analytical_2}).}
        (b) Variation of the diffusiophoretic velocity (at $x=0.5$) with time for varying $\beta$ (for solute-out mode).
        (c,d) Temporal evolution of effectiveness for withdrawal (c) and injection (d). Solute-in (dashed line) and solute-out (solid line) modes are shown for various magnitudes of applied solute gradients at $D_s=10^{-9}$ m$^2$/s, $D_p=10^{-12}$ m$^2$/s, and $\Gamma_p=10^{-9}$ m$^2$/s.
        \rev{Here, (c) and (d) share the same legends for $\beta$.} 
        }
\label{fig:performance_measure}
	\end{figure*}

To discern the difference in rapidness in two modes, we first note that the solute-out mode offers larger magnitudes of velocity due to the logarithmic nature of $\IB{u}_p \sim \nabla c/c$; higher the local concentration (relative to the local gradient), lower the velocity magnitude.	
The distinct spatial pattern of $n$-profile  can be interpreted from the diffusiophoretic velocity profiles for the two modes. Although $\IB{u}_p$ monotonically decays in time for both modes, there are notable differences with respect to spatial variation (depicted in Fig.\ref{fig:Vel_cz}(a)), which can be understood by recalling that diffusiophoretic velocity is ratio of local solute gradient to local solute concentration. The former always decreases away from pore mouth, the latter is different in the two modes: (\textit{i}) for solute-out modes, the solute concentration increases with pore length, resulting in both factors contributing to a monotonic decay with $x$; (\textit{ii}) whereas for solute-in mode, the solute concentration decreases more rapidly than the decrease in gradient, yielding in a maximum in the middle of the channel.
These disparate velocity profiles of the two modes bring in the qualitative changes in the two $n$-profiles. 
Furthermore, the monotonic decay of $\IB{u}_p$ with $x$ (for solute-out) yields in a faster withdrawal of the particles that are closer to the pore mouth, i.e., `shallow' withdrawal. 
 For solute-in case, the maximum withdrawal velocity in the middle of the pore draws particles in from the deep-end, which are released (relatively slowly) from the pore mouth, yielding in a `deeper' withdrawal accompanied by a pore accumulation.

\textit{Injection.}
Figure \ref{fig:with_inj} (a) shows the injection profiles of bare colloidal diffusion, i.e., when no external electrolyte gradient is applied: the colloids slowly diffuse inside the pore over $\sim 100 t_d$. Fig. \ref{fig:with_inj} (d) shows that both modes can inject substantial amount of colloids in $\sim t_d$ times, a two-order-of-magnitude enhancement. 
Furthermore, it can be observed that while solute-out mode injects high density of colloids with propagating fronts, solute-in mode injection is relatively uniform.
These fronts signify a boundary layer region $\delta(x)$ where (otherwise weak) colloid diffusion balances diffusiophoretic advection. 
This balance reveals that the boundary layer grows as $\delta \sim D_p \, x/\Gamma_p $ \footnote[4]{To extract the characteristic scales from the governing equation (\ref{n-GE}), we note that the balance at time $t$ is: $\frac{D_p}{\delta^2} \sim \frac{\IB{u}_{p\,ch}}{\delta}$, where $\IB{u}_{p\,ch} \sim \Gamma_p/x$ is the scale for location-dependent diffusiophoretic magnitude.}.
We also underscore a subtle aspect that, at early times ($t<t_d$), the solute-in mode enables a deeper injection. However, this is not always the case, as we later demonstrate that at lower mobility, the solute-out mode remains dominant at all times.

To understand these distinct patterns of injection in two modes, we turn our attention back to Fig.\ref{fig:Vel_cz}.
For solute-out mode, the monotonic spatial decay of $\IB{u}_p$ suggests that the colloids enter the pore at high velocities and result in accumulated bands or fronts. These fronts then gradually propagate and broaden within the pore as their constituent colloids continue to experience the logarithmic solute gradients. 
On the other hand, for solute-in mode, $\IB{u}_p$'s non-monotonic decay suggests that colloids experience a gentle rise in injection velocity with $x$ (contrary to solute-out), resulting in a gradual dilution of colloidal density. 
Fig. \ref{fig:Vel_cz}(b) shows how stronger solute gradients yield in higher magnitude of phoretic velocity for both modes.

\subsubsection*{Persistence time}
\noindent
 Given these differences between the two modes, it becomes evident that the lifespan of phoretic transport cannot be adequately characterized by $t_d$.
 Hence, we define persistence time that is calculated as the time taken for $ \IB{u}_p$ to decay to 99\% \rev{from its maximum value}, averaged over the entire pore length: $\langle \tau \rangle = \frac{1}{L} \int_{0}^{1} \tau(x), \, \text{where } \tau$ the local persistence is defined such that  $\IB{u}_p(x,\tau(x)) = 0.01 \, \IB{u}_p(x,\rev{0.1}) $. 
 \rev{In our calculations, we consider earliest time to be $t^*=0.1 t_d$.}
This definition gives us an average measure of lifetime of diffusiophoretic process and is identical for injection and withdrawal.
Figure \ref{fig:performance_measure} (a) shows its variation with the applied concentration difference, a ratio characterized by $\beta$ whose variation spans for both solute-out ($\beta<1$) and solute-in modes ($\beta>1$).
We note that solute-out diffusiophoretic modes are relatively more persistent and are quite sensitive to the externally applied gradient:  $\tau$ rises logarithmically with diminishing $\beta$. 
This can be directly deduced by simplifying the phoretic velocity for times \( t \gtrsim t_d \), enabling a one-term approximation in the infinite series solution for the solute profile, which yields:
\begin{equation}
    \IB{u}_p = \frac{\Gamma_p \cos(\pi x/{2})}{e^{\lambda t}(\frac{\beta}{1-\beta})+\frac{4}{\pi} \sin(\frac{\pi x}{2})},
\end{equation}
where $\lambda = \pi^2/4$. This translates to the following conditions for $\tau$ in the two extremes:
\begin{gather}\label{analytical}
\frac{\IB{u}_p}{\IB{u}_p\left. \right \vert_{(t\rev{\to}0)}} =
\begin{cases}
\displaystyle
\frac{\frac{4}{\pi} \sin\left(\frac{\pi x}{2}\right)}
     {\frac{4}{\pi} \sin\left(\frac{\pi x}{2}\right) + \beta e^{\lambda t}}  
\sim  \frac{1}{1+ \beta e^{\lambda t}}  
\; \text{for}  \; \beta \ll 1   \\[12pt]
\displaystyle
\frac{\frac{4}{\pi} \sin\left(\frac{\pi x}{2}\right) - 1}
     {\frac{4}{\pi} \sin\left(\frac{\pi x}{2}\right) - e^{\lambda t}}  
\sim  \frac{1}{e^{\lambda t}}  
\; \text{for} \; \beta \gg 1
\end{cases}
\end{gather}
The persistence of extreme solute-in mode ($\beta \gg 1$) shows independence with respect to the magnitude of applied solute gradient, whereas the extreme solute-out mode ($\beta \ll 1$) demonstrates a delayed exponential decay that scales with $\sim \ln \left[ \beta^{-1/\lambda} \right]$, where $1/\lambda=4/\pi^2 \approx 0.4$. The latter is also explicitly shown in Fig. \ref{fig:performance_measure} (b), where a reduction in $\beta$ delays the exponential decay of phoretic velocity. We note that, the only condition that can  sustain the diffusiophoretic process indefinitely is an environment absolutely free of solute ($\beta = 0$) at all times. 
\rev{However, we note that this ideal case cannot be experimentally realized as, at extremely low solute concentrations,  the continuum approximation breaks down, as the distance between the ions can become comparable or larger than particle size.
}

\rev{
The persistence time can also be analytically estimated within the one-term approximation. Plugging the aforementioned $\IB{u}_p$ into $\frac{\IB{u}_p}{\IB{u}_{p}\vert_{t=0.1}}=0.01$, approximately yields the following equation for local persistence time:
\begin{equation}
\label{analytical_2}
    \tau = 0.405 \log\left[128 + 126\left(\frac{1-\beta}{\beta}\right) \sin(\pi x/2)\right].
\end{equation}
This expression can be integrated over the non-dimensional pore length to yield the average persistence $\langle \tau \rangle$, which agrees well with numerical calculation and asymptotic estimates in Fig.4 (a) as empty circles.}

\subsubsection*{Effectiveness}
Persistence time characterization aids in concluding that solute-out mode has relatively larger lifespan. 
\rev{As an additional performance metric, we evaluate \textit{effectiveness}, which, for the case of withdrawal, is the ratio of the total colloids withdrawn in the presence of solute gradients relative to those removed by bare colloidal diffusion (in the absence of solute gradient, i.e., $\beta=1$). For injection, it is defined analogously as the ratio of the total colloids injected in the presence of solute gradients to those injected by pure colloidal diffusion ($\beta=1$).
}
\begin{align}
\mathcal{E}_{with} (t) &= \frac{1- \int_{0}^{L} n(x,t) \, dx}{1- \int_{0}^{L} n_{\beta=1}(x,t) \, dx}, \\
\mathcal{E}_{inj} (t) &= \frac{ \int_{0}^{L} n(x,t) \, dx}{ \int_{0}^{L} n_{\beta=1}(x,t) \, dx}.
    \label{eq:eff_with}
\end{align}
Here $n$ is obtained from solving the coupled system of equations (\ref{c-GE},\ref{n-GE}). Figures \ref{fig:performance_measure} (c,d) show the temporal decay in the effectiveness of phoretic transport. 
For both modes, we note an order of magnitude enhancement compared to bare colloid diffusion, which persists for several $t_d$. 
Furthermore, while the solute-out mode is generally more effective, the case of injection is further improved because the ambient environment is an infinite reservoir of colloids, whereas for withdrawal, pore's capacity is finite.
For the same reason, in Fig. \ref{fig:performance_measure}(c) we observe that effectiveness for withdrawal saturates at higher magnitudes of solute-in/solute-out gradients.

\rev{The mobility model considered here is appropriate for micron-sized colloids, where the Debye layer is extremely thin compared to the particle size. For submicron or nanoscale colloids, however, the Debye layer has a finite thickness that varies with concentration, rendering the mobility a function of the local solute concentration, which evolves both spatially and temporally \cite{keh2000diffusiophoretic,ganguly2023diffusiophoresis}. In the Appendix, we incorporate this mobility model and show that while finite Debye layer effects introduce quantitative modifications—such as dampening of colloidal withdrawal and injection—they do not affect the qualitative conclusions presented here.}


\begin{figure*}[t]
		\centering
        {\includegraphics[width=1.02\linewidth]{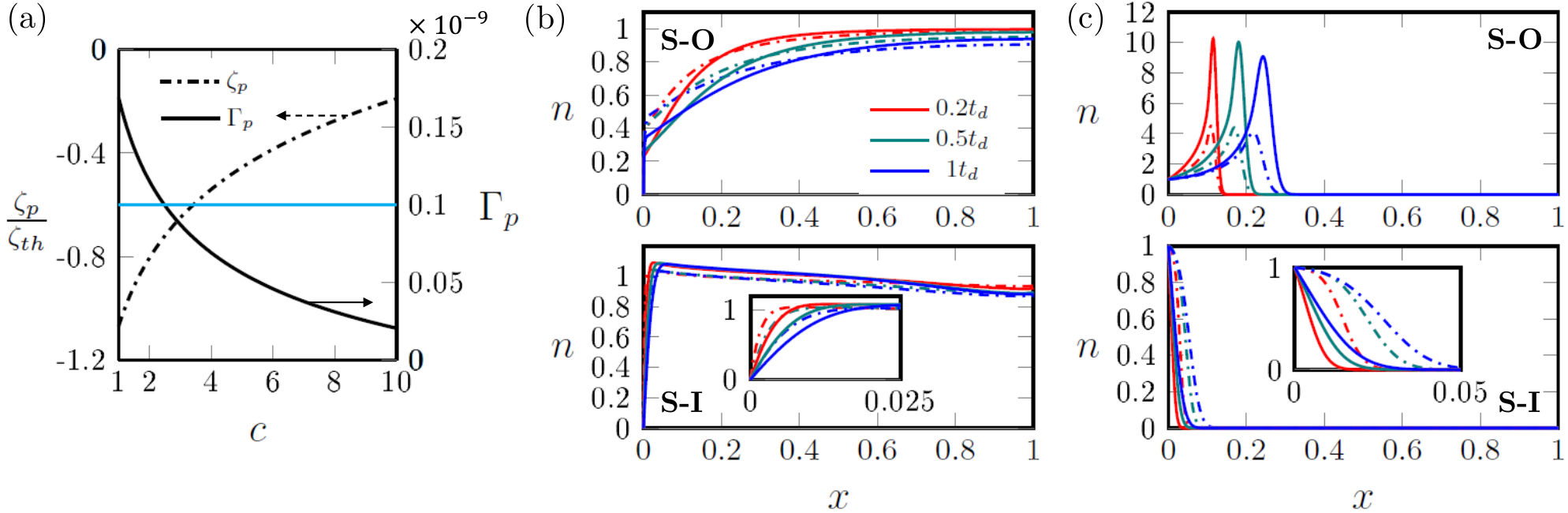}}
	\caption{\textbf{Colloidal transport for constant charge model}.
    (a) The variation of particle zeta potential (scaled by $k_B T/ze$) and mobility ($m^2/s$) variation with solute concentration. 
  The solid line at $\Gamma_p = 0.1 \times 10^{-9}$ $m^2/s$ depicts the value used for constant mobility comparison.
    Colloid concentration profiles for (b) withdrawal and (c) injection. 
    Top figures depict solute-out mode, bottom figures correspond to solute-in.
    Dash-dotted curves correspond to constant mobility results. 
    Parameters: $D_s=1.61\times10^{-9}$ m$^2$/s, $D_p=4\times10^{-13}$ m$^2$/s, $\Gamma_p=10^{-10}$ m$^2$/s (constant zeta potential) and $\Gamma_p=f(c)$ (constant charge model Eq. \ref{eq:GammaP}).
We note that the propagating fronts of the colloid density shrink over time because the mobility is lower on average than the previous case (c.f. Fig.\ref{fig:with_inj}d).    }
    \label{fig:with_inj_cz_cc}
	\end{figure*}

\subsection*{Withdrawal and injection of colloids with variable mobility}
\noindent 
We now extend our analysis by relaxing the assumption of constant mobility.
Specifically, colloids' zeta potential is the function of the local solute concentration, influencing the concentration-dependent mobility, i.e., the mobility experiences spatio-temporal variation.
\rev{This variability in mobility arises from counterion shielding of surface charges (associated with changes in the Debye layer thickness), resulting in a concentration-dependent modification of the zeta potential \cite{kirby2004zeta}.}
This `constant charge model' assumes that the surface charge density remains fixed and for base zeta potentials larger than thermal potential, its modification is logarithmic $\left( \sim a+b\log(c)  \right)$  \citep[eq.8]{kirby2004zeta}, which would consequently alter the mobility $\Gamma_\text{p}(c)$ as:
\begin{equation}
	 \frac{\varepsilon}{2\mu}\left(\frac{k_\text{B}T}{\rev{Z}e}\right)^2 \left[2\beta_s\frac{Ze\zeta_p (c)}{k_BT} + 8 \ln cosh\left(\frac{Ze\zeta_p (c)}{4k_\text{B}T}\right)\right],
	\label{eq:GammaP}
\end{equation}
where $\beta_s(=\frac{\mathcal{D}_{+}-\mathcal{D}_{-}}{\mathcal{D}_{+}+\mathcal{D}_{-}})$, $\varepsilon$, $\mu$, $k_B$, $\rev{Z}$, $e$, and $T$ represent the diffusivity contrast between cation and anion, medium permittivity, medium viscosity, Boltzmann constant, ionic valencies, elementary charge, and absolute temperature, respectively. 
\rev{The expression for the zeta potential is motivated by the study of \citet{akdeniz2023diffusiophoresis} on size-controlled PS-PEG particles. They found that $\zeta_p = a + b \log_{10}(c^\ast/1\text{mM})$, where $a = -27.43$ mV, $b = 22.63$ mV, and $c^\ast$ ranges from 1 to 10 mM. This expression is valid in the limit where the thickness of the electrical double layer is negligible compared to the particle size—an approximation that holds for the micron-sized particles used in \citet{akdeniz2023diffusiophoresis}.
}
In our analysis below, we compare the modifications to colloidal density profile that are caused by concentration-dependent mobility. 

     \begin{figure}[h]
		\centering
        {\includegraphics[width=1\linewidth]{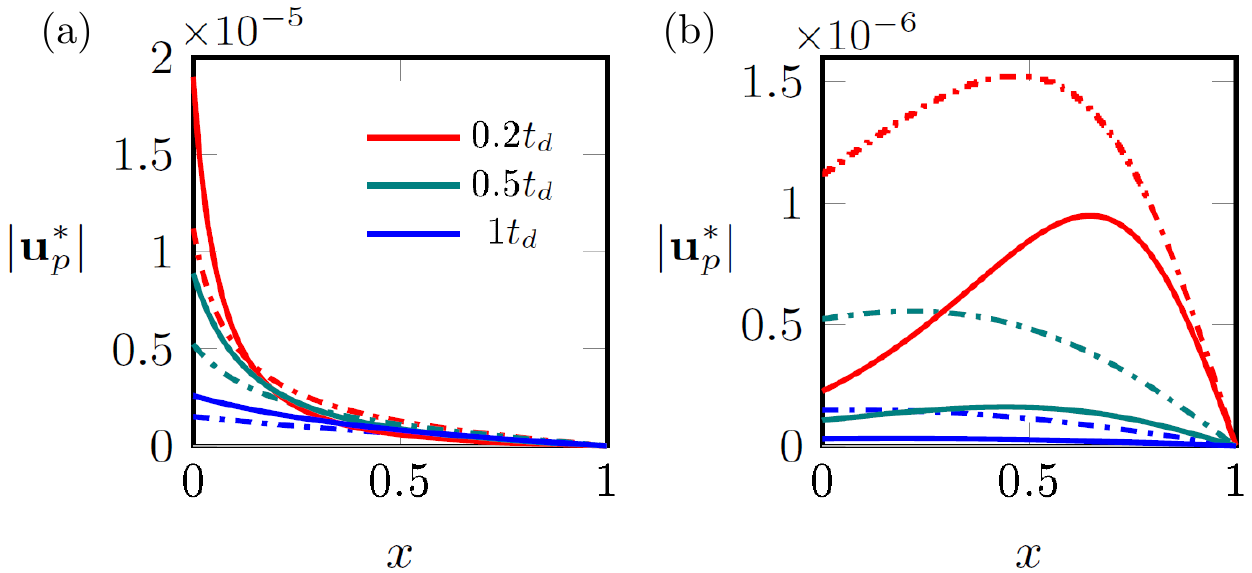}}
	\caption{ \textbf{Phoretic velocity for variable mobility model.}
    Comparison of particle velocity (m/s) (a) solute-out and (b) solute-in modes for constant zeta (dash-dotted line) and constant charge (solid line) models at $D_s=1.61\times10^{-9}$ m$^2$/s, $\Gamma_p=10^{-10}$ m$^2$/s (constant zeta potential) and $\Gamma_p=f(c)$ (constant charge model Eq. \ref{eq:GammaP}).}
    \label{fig:Vp_cz_and_cc}
	\end{figure}

        \begin{figure*}[t]
		\centering
         {\includegraphics[width=0.85\linewidth]{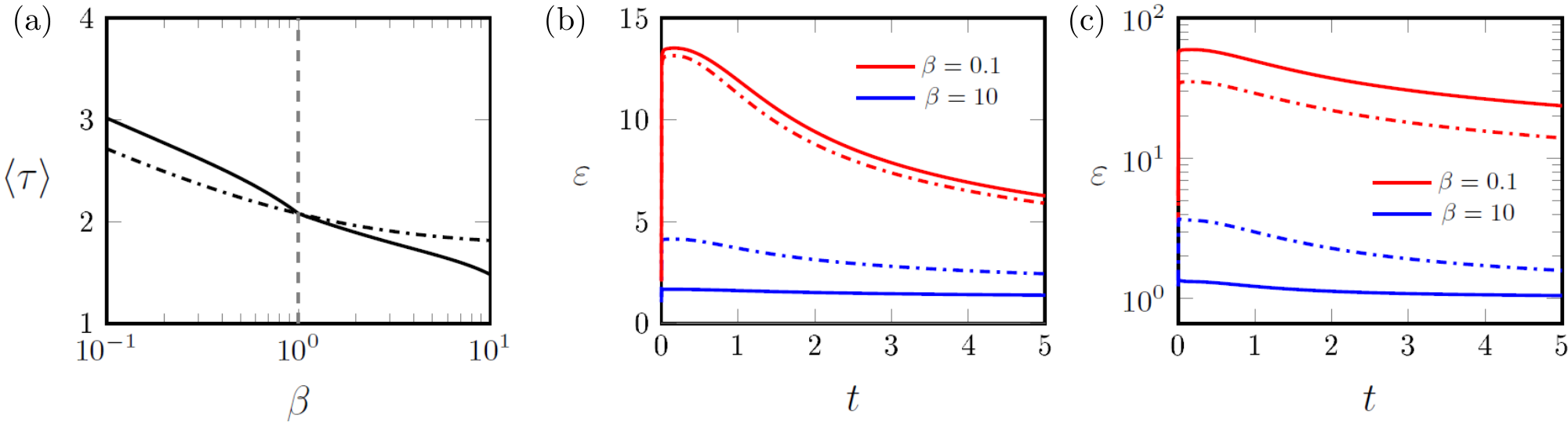}}
	\caption{\textbf{Performance measures for constant charge model.}
    (a) Variation of persistence time with applied concentration difference ratio $\beta$. 
    Dash-dotted curve corresponds to constant mobility model.
    Effectiveness of (b) withdrawal and (c) injection cases for different solute gradient modes as solute out (red) vs solute in (blue) at $D_s=1.61\times 10^{-9}$ m$^2$/s, $D_p=4\times 10^{-13}$ m$^2$/s, $\Gamma_p=10^{-10}$ m$^2$/s (constant zeta potential) and $\Gamma_p=f(c)$ (constant charge model). 
    }
    \label{fig:cc_mode_comparison}
	\end{figure*}

    \begin{figure*}[t]
		\centering
{\includegraphics[width=1\linewidth]{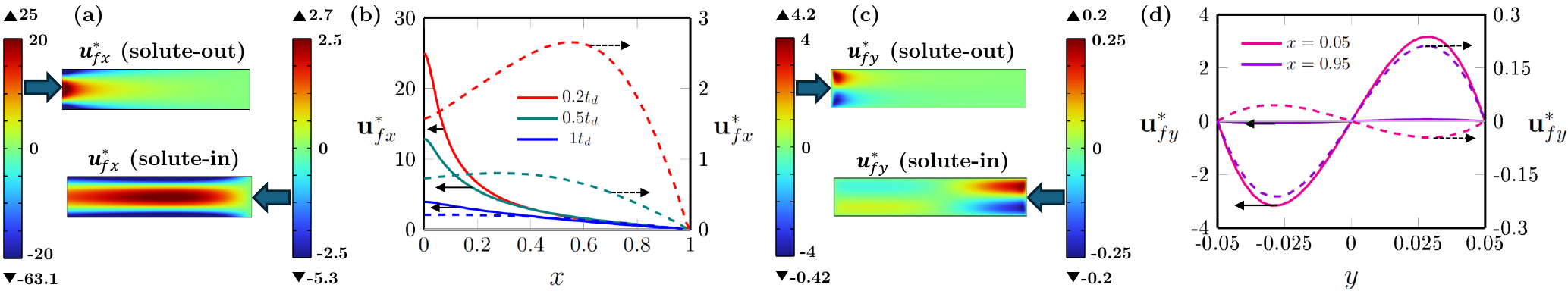}}
		\caption{\textbf{Osmotic flow profile.} (a,b) Contours and centerline plot for x-component flow velocity ($\mathbf{u}^\ast_{fx}$, $\mu$m/s), respectively. The arrowbars indicate the colorbar corresponding to each contour.
        (c,d) Contours and vertical variation of the y-component osmotic velocity ($\mathbf{u}^\ast_{fy}$, $\mu$m/s). 
         The solid and dashed lines represent the solute-out for $\Gamma_w>0$ and solute-in for $\Gamma_w<0$ modes, respectively. 
          The contours are for $t=0.2t_d$. Parameters: {$D_s=1.61\times 10^{-9}$ $m^2$/s and} $\Gamma_w=g(c)$. Variation of mobility shown in \S2 in Supplementary material.} 	\label{fig:DO_vel}
\end{figure*}

Figure \ref{fig:with_inj_cz_cc}(a) shows the extent of this variation, where the dash-dotted line represents constant zeta-potential value that we have chosen for reference.
This mobility decay (with increasing local concentration) corresponds to the PS-PEG particle ($\Gamma_p>0$) which will exhibit injection for solute-out mode and withdrawal for solute-in mode. 
To study the two other modes of solute-in and solute-out transport, we also consider an equivalent particle with mobility of opposite sign, which could be realized by using NaOH {($\beta_s = {-0.596} $)} as solute instead of NaCl ($\beta_s = {-0.208} $).

Fig. \ref{fig:with_inj_cz_cc}(b) shows that withdrawal dynamics is marginally impacted by the incorporation of the constant charge model. For the solute-out mode, colloid withdrawal is enhanced closer to pore entry, whereas for the solute-in mode, withdrawal is hampered by the high solute concentration.
On the other hand for injection, in Fig.\ref{fig:with_inj_cz_cc} (c), both modes are significantly influenced as the transport of colloids is dictated by dynamics at the pore entry. Here, the solute-out mode offers lower concentrations, whereas the solute-in mode dampens mobility at this crucial junction due to the high local concentration. 
Overall, Figs.\ \ref{fig:with_inj_cz_cc}(b,c) show that for the constant charge model, disparity in the qualitative and quantitative dynamics of the solute-in/out modes has widened relative to that shown in Fig. \ref{fig:with_inj}. This is also consistent with the trends driven by amplification and dampening in $\IB{u}_p$ (shown in Fig.\ref{fig:Vp_cz_and_cc}(a) and (b), respectively).

In the context of performance, the persistence time distribution is also altered and loosely depicts a linear variation in Fig.\ref{fig:cc_mode_comparison}(a): unlike the constant zeta model, here the lifespan of phoresis in solute-in mode is sensitive and decays with $\beta$, as the high concentrations dampen both the mobility and logarithmic gradient.
For effectiveness, the temporal trends for injection and withdrawal are shown in Fig. \ref{fig:cc_mode_comparison}(b, c), these trends are consistent with those shown in Fig. \ref{fig:with_inj_cz_cc}(b, c): the solute-out mode exhibits enhanced effectiveness due to amplified mobility near the pore mouth, whereas the solute-in mode is ineffective due to damped mobility. 
The stark difference in the magnitude of modification between injection and withdrawal (compare y-axes) arises from the fact that, during injection, the enhancement in $\IB{u}_p$ can transport colloids from a reservoir with abundant colloids, whereas during withdrawal, the colloid supply is finite from the pore.

\subsection*{Influence of wall-induced osmotic flow}
    \begin{figure*}[t]
		\centering
{\includegraphics[width=0.95\linewidth]{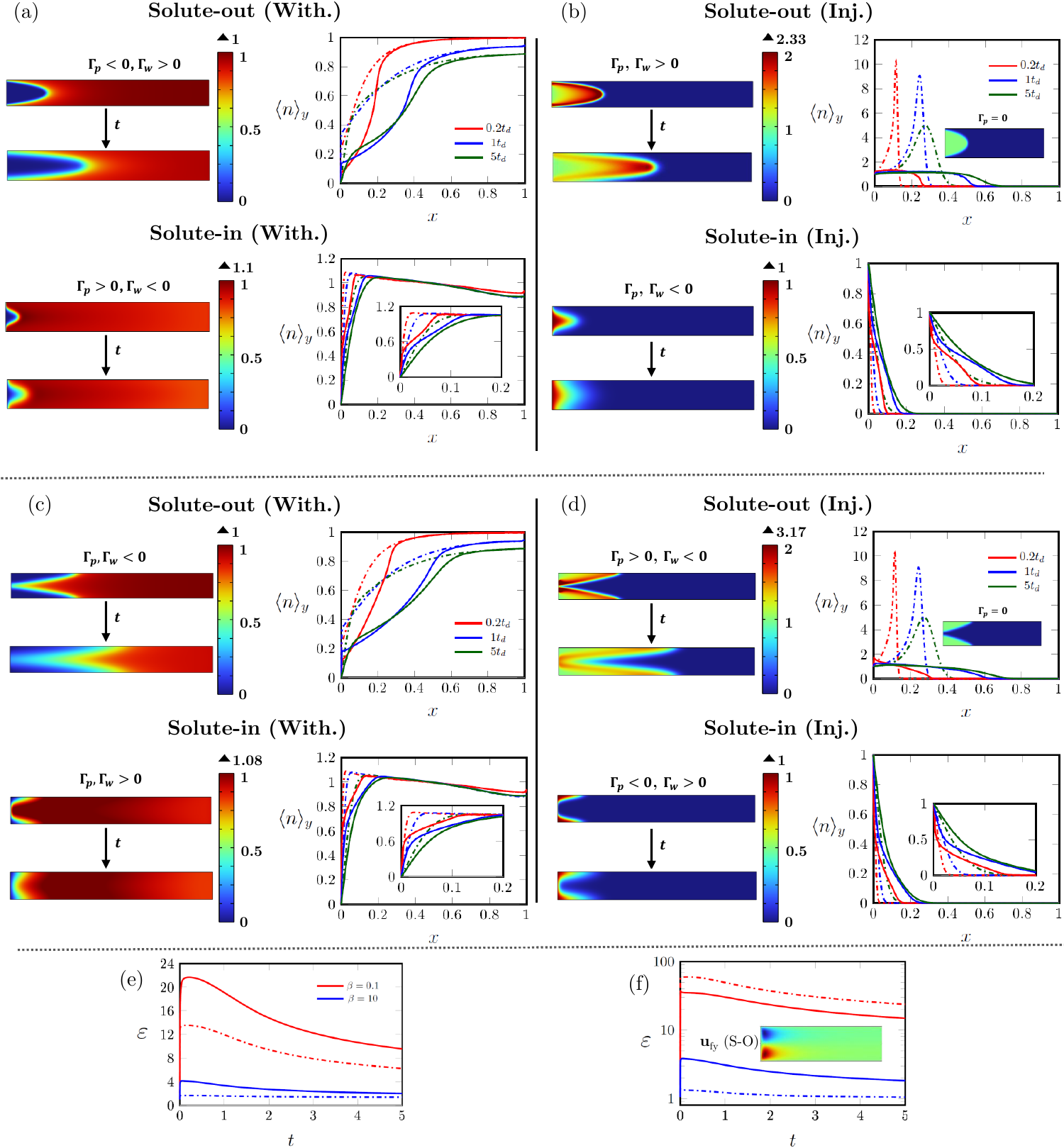}}
	\caption{\textbf{Spatiotemporal maps with osmotic flow interference}. (a) Withdrawal and (b) injection contours of colloids at $0.2 \, t_d$ and $1 \, t_d$. 
    Line plots show variation of y-averaged concentration profile $\langle n \rangle {}_y$ with osmotic flow ($\Gamma_p=f(c)$, $\Gamma_w=g(c)$; solid line) and without osmotic flow ($\Gamma_w=0$; dot-dashed line). The inset in solute-out plot show the absence of particle banding when phoretic effects are absent.
    (c,d) Spatiotemporal patterns for other combinations of wall and particle mobilities for withdrawal and injection, respectively.
    (e,f) Effectiveness of withdrawal (a) and  injection (b) cases for solute out (red) and solute in (blue) modes, respectively. Dot-dashed curves correspond to results for $\Gamma_w=0$. Inset shows the intensity of lateral flow velocity near the pore mouth, which tends to push the particles towards the centerline.
    Parameters: $D_s=1.61\times10^{-9}$ m$^2$/s and $D_p=4\times10^{-13}$ m$^2$/s.  }
    \label{fig:with_inj_cc_DO}
	\end{figure*}

\noindent
Here we consider the cases when walls exhibit significant mobility such that, in the presence of external solute gradient, a diffusio-osmotic slip is generated \cite{anderson1989} ($\IB{u}_\text{slip} = - \frac{\Gamma_w(c)}{D_s} \nabla \log c $). This slip-driven flow further renders a bulk motion in the pore and has been shown to affect particle dispersion \cite{shin2016size,kar2015enhanced}. It has also been utilized to trap colloids and vesicles in bioassays \cite{rasmussen2020size}, as well as for performing low-cost zeta potentiometry \cite{shin2017low,ault2019characterization}.  
To begin the analysis, we first note from Eq.(\ref{n-GE}) that osmotic flow would not contribute significantly to the colloidal dispersion in $\text{Pe} \ll \Gamma_p/D_s $ regime. Hence in the discussion below, we consider the wall mobilities that yield osmotic flows in $\text{Pe} \sim \Gamma_p/D_s $ regime. The chosen parameters yield in osmotic and phoretic velocities that broadly match with the experiments of \citet{kar2015enhanced}.

Figures. \ref{fig:DO_vel} (a,b) illustrate the horizontal component of osmotic flow that occurs for the two solute-gradient modes. 
To render the shape of bi-directional osmotic flows similar for the two modes (ensuring an equitable comparison), the wall mobilities ($\Gamma_w$) are set such that positive values are assigned for the solute-out mode and negative values for the solute-in mode. Other combinations would yield a bi-directional flow of inverted shape. 
In Fig. \ref{fig:DO_vel}(a,b), the wall mobility senses the chemical gradients and induces a strong outward wall slip, and as the net flow rate within the pore is zero, a parabolic inflow develops away from the walls \cite{ault2019characterization}.
For solute-out mode, this bidirectional flow decays monotonically along the pore depth, whereas non-monotonically for solute-in mode; a trend similar to diffusiophoretic velocities shown in \ref{fig:Vp_cz_and_cc}(a,b).
Both of these modes generate a modest convective flow ($\sim \Gamma_p/D_s$) as shown by the temporal decay of the spatially-averaged flow in \S3 in the Supplementary material.
Figs. \ref{fig:DO_vel} (a,b) depict the vertical component of the osmotic flow, which is responsible for the lateral redistribution of colloids. This flow arises due to the dead-end nature of the pore; since the net flow rate is zero, the slip-driven motion of the incompressible fluid (\(\partial \IB{u}_{fx}/\partial x \neq 0\)) induces a nonzero \(\IB{u}_{fy}\) component.
For the chosen signs of wall mobilities (or surface potentials), \(\IB{u}_{fy}\) directs the colloids away from the centerline. Interestingly, while recirculation occurs near the pore entry in the solute-out mode (as observed here and reported in ref. \cite{shin2016size,alessio2021diffusiophoresis}), the solute-in mode instead exhibits recirculation at the end of the pore. This is because the osmotic slip is sustained throughout the pore length and \(\partial \IB{u}_{fx}/\partial x \) is largest near $x^*=L$, and thus $\IB{u}_{fy}$ is largest here. 
This difference in recirculation location for the two modes will be shown to cause significant deviations in transport efficacy.
Finally, we note that these trends in slip and bulk velocity are exactly reversed between the solute-out and solute-in modes when the mobilities are switched.
For completeness, the impact of both combinations of zeta potentials is considered in Fig. \ref{fig:with_inj_cc_DO}, which demonstrates the interference of osmotic flow on spatiotemporal patterns of colloids. Note that both combinations yield near-identical effectiveness results.

Fig. \ref{fig:with_inj_cc_DO} (a) shows the results for colloidal withdrawal for two modes (in the top and bottom rows). 
Here, the osmotic flow is configured such that an outward wall slip ejects particles into the ambient fluid, which is accompanied by an inward bulk flow. 
The contours are supplemented by the plots that quantify the impact on $y$-averaged profiles.
The solute-out mode, which is more responsive to chemical gradients, is substantially influenced by osmotic flow interference, also reflected in the effectiveness measurement shown in Fig. \ref{fig:with_inj_cc_DO}(e).
For injection, Fig. \ref{fig:with_inj_cc_DO} (b) illustrates similar patterns in concentration profiles, with certain intriguing features that influence both the y-averaged profile and overall effectiveness, a closer look at which follows.
The outward osmotic slip at the walls competes with the inward phoretic injection, leading to local colloidal accumulation near the pore entry corners. Whereas, the inward osmotic flow in the channel bulk convects the injecting particles deeper, consequently shaping the injecting band of colloids as a parabola \cite{shin2016size}.
In agreement with the observations of \citet{shin2017low} experiments, we find that this banding is further exacerbated by $\IB{u}_{fy}$ convection, which is strongest near the pore for solute-out and, as shown in Fig. \ref{fig:DO_vel}(c), acts to push particles toward the outward flow near the walls.
As a result of this lateral migration, the effectiveness of the solute-out mode suffers substantially from the osmotic interference---a trend contrary to that of withdrawal.

The results for other combinations of wall mobilities [Fig. \ref{fig:with_inj_cc_DO}(c,d)] exhibit similar quantitative trends, with $y$-averaged colloidal concentrations being near-identical to previous case.
The spatiotemporal profile, however, now resembles a reverse concentration boundary layer at the top and bottom walls due to the bidirectional flow. 
Interestingly, the injection profile in Fig. \ref{fig:with_inj_cc_DO} (d) shows inverse banding.
This is because, for these wall mobilities, the solute-out mode induces inwawrd osmotic slip and outward bulk flow; the latter competes with the inward phoretic injection, leading to local colloidal accumulation near the central axis of the pore entry. This banding is further shaped by $\IB{u}_{fy}$ convection [shown in Fig. \ref{fig:with_inj_cc_DO}(f) inset]. This lateral convection is strongest near the pore for solute-out and, as shown in the inset contour, acts to push particles toward the outward flow near the centerline, which consequently affects the injection efficiency as shown in Fig. \ref{fig:with_inj_cc_DO}(f). 
\rev{
Finally, we note that wall-induced diffusioosmosis contributes to the spreading of the colloidal distribution, as evidenced by the \(y\)-averaged profiles for injections with and without wall slip (solid and dashed curves, respectively) shown in Fig.~9(b,d). Additional results from \S 4 of the Supplementary Material demonstrate that the extend of spread rendered by the wall mobility of two types ($\Gamma_w>0$ and $\Gamma_w<0$), consistent with earlier findings~\cite{alessio2021diffusiophoresis,alessio2022diffusioosmosis}.
}

%
%

\section*{Conclusion}

\noindent
Aiming to contribute to a deeper understanding of diffusiophoretic (DP) transport within electrolytic gradients in complex media, the current work demonstrates the significant role of gradient orientation in shaping colloidal migration within dead-end pores. 
This study focuses on withdrawal and injection dynamics, with key findings presented below in relation to the questions outlined in the Introduction section.
    
    1. We find that the solute-out mode facilitates rapid yet shallower withdrawal, whereas solute-in mode enables deeper removal of colloids from the pores. For injection, colloids under the solute-out (S-O) mode propagate rapidly as accumulated non-uniform bands, while the solute-in (S-I) mode drives a notably uniform and gradual transport.
    The difference in rapidness for the two modes has to do with difference in local magnitude of concentration, as the transport is driven by $\nabla c/c$. The spatial distinction in the patterns has to do with the difference in the DP velocity profiles: for S-O, it decays sharply and monotonically in space (enabling colloidal accumulation), whereas the non-monotonic profile for S-I facilitates a maximum well within the pore (ensuring uniformity).
    
   2. Given the differences in spatiotemporal evolution, the lifespan of injection/withdrawal cannot be characterized by the solute-diffusion timescale. 
    Hence, we introduce a spatially-averaged persistence time that shows: (\textit{i}) for strong S-O gradients ($\beta \ll 1$), transport decays with a delayed exponential dependence and persistence scales as  $\ln \left( \beta^{-0.4} \right)$; (\textit{ii}) for strong S-I gradients ($\beta \gg 1$), the transport is independent of the applied gradient. Overall, solute-out mode has relatively higher persistence.
    Furthermore, both modes exhibit an order of magnitude enhancement in effectiveness, with S-O being the more pronounced mode.
    For higher magnitudes of mobility ($\Gamma_p \gtrsim 10^{-9} m^2/s$), S-I's transport can compete with S-O mode. Furthermore, we find that variable phoretic mobility (incorporated via the constant charge model for large zeta potentials) can further exacerbate the differences between the two modes due to the added effect of concentration-induced mobility damping.
    
    3. The diffusio-osmotic flow, that occurs due to finite wall mobilities, has a bidirectional signature because the net flow rate inside the dead-end pore is zero. 
    S-O generates a strong axial flow near pore entry that monotonically decays with $x$, accompanied by significant lateral flow components. On the other hand, S-I induces a modest axial flow throughout the pore with a maximum well within it, resulting in its strongest lateral flow components located at the pore end.
    We further studied the impact of these flows on the colloidal distribution: the osmotic flows of S-I overall enhance both withdrawal and injection, however, the effectiveness of injection suffers for S-O mode. This occurs due to near-pore lateral convection, which directs particles toward the outward stream of the bidirectional osmotic flow. These results are validated across all four combinations of mobility parameters, for both withdrawal and injection.

The current results can aid in a deeper understanding of particle and drug transport in porous biofilms  \cite{somasundar2023diffusiophoretic} and provide insights into the injection and withdrawal of non-motile microorganisms, recently shown to be influenced by surfactant gradients \cite{doan2020trace}. 
Additionally, while the often-overlooked solute-in mode exhibits slower colloidal injection and withdrawal, its distinct characteristics may be advantageous in applications requiring uniform dispersion and deep penetration \cite{velegol2016origins}. For example, this mechanism could be beneficial in shear-thickening fabric manufacturing, where rapid and homogeneous particle infusion into fabric pores can enhance mechanical properties such as tensile strength and durability \cite{he2018impact,talreja2017functionalization}.

\rev{
\citet{banerjee2019long} devised hydrogel-based `soluto-inertial' source and sink beacons that facilitate spatio-temporal control over colloidal suspensions. Later, for non-electrolytic gradients, \citet{raj2023two} modeled 2D colloidal banding and reported optimal designs of such beacons that exhibit time-dependent solute emission/absorption.
Recently, \citet{alessio2023diffusiophoresis} modeled pigment cells (chromatophores) as non-electrolytic diffusiophoretic particles and demonstrated that, in the presence of two chromatophore types and two solute species, augmented Turing patterns can emerge with significantly finer length scales.
Based on the insights from \S III on electrolytic gradients, we hypothesize that colloids with positive mobility will exhibit large and rapid wavefronts when attracted to regions of higher solute concentration, whereas colloids with negative mobility will form slower and uniform patterns. 
This distinction in pattern morphology and timescales may have important implications for the optimal design of soluto-inertial beacons, colloidal banding phenomena, and the understanding of cellular pattern formation, highlighting the need for further investigation.
}

\renewcommand{\thefigure}{A.\arabic{figure}}
\setcounter{figure}{0}

\begin{figure*}[t]
    \centering
    {\includegraphics[width=0.8\linewidth]{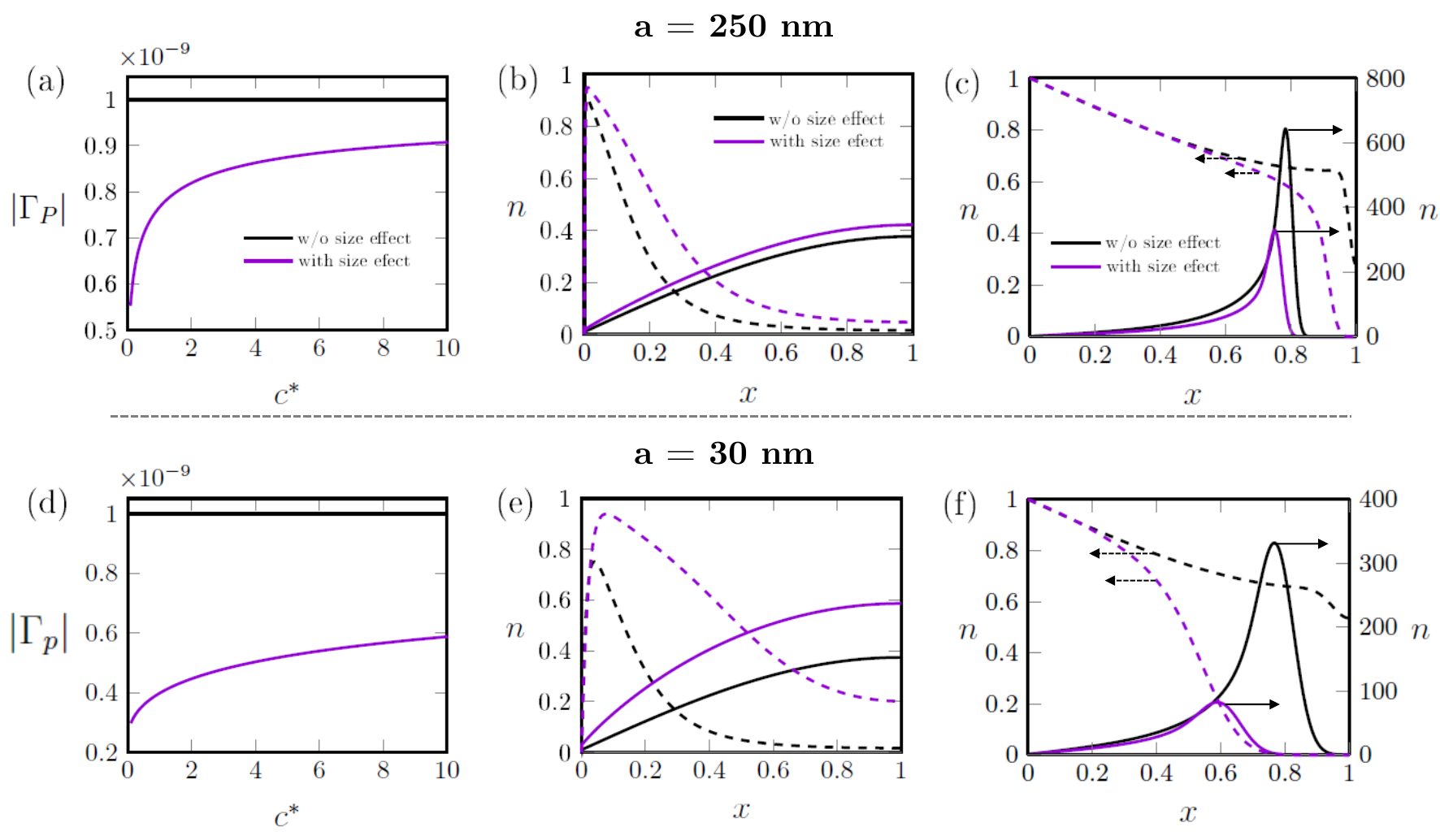}}
     \caption{(a) and (d) Particle mobility ($\Gamma_p$, m$^2$/s) variation with solute concentration ($c^\ast$, mM) inside the pore with and without Debye layer effect for $a=250$ nm and $30$ nm, respectively. (b,c) Particle withdrawal and (e,f) injection patterns for solute-out (solid line) and solute-in mode (dashed line) at $a=250$ nm and $30$ nm, respectively. Parameters: $D_s=1\times 10^{-9}$ m$^2$/s, $t=0.5$, $D_p=1\times 10^{-12}$ m$^2$/s ($a=250$ nm) and $8.084\times 10^{-12}$ m$^2$/s ($a=30$ nm), $\beta=0.01$ (S-O) and $100$ (S-I).}
    \label{fig:E1}
    \end{figure*}
 \begin{figure*}[t]
    \centering
    {\includegraphics[width=0.85\linewidth]{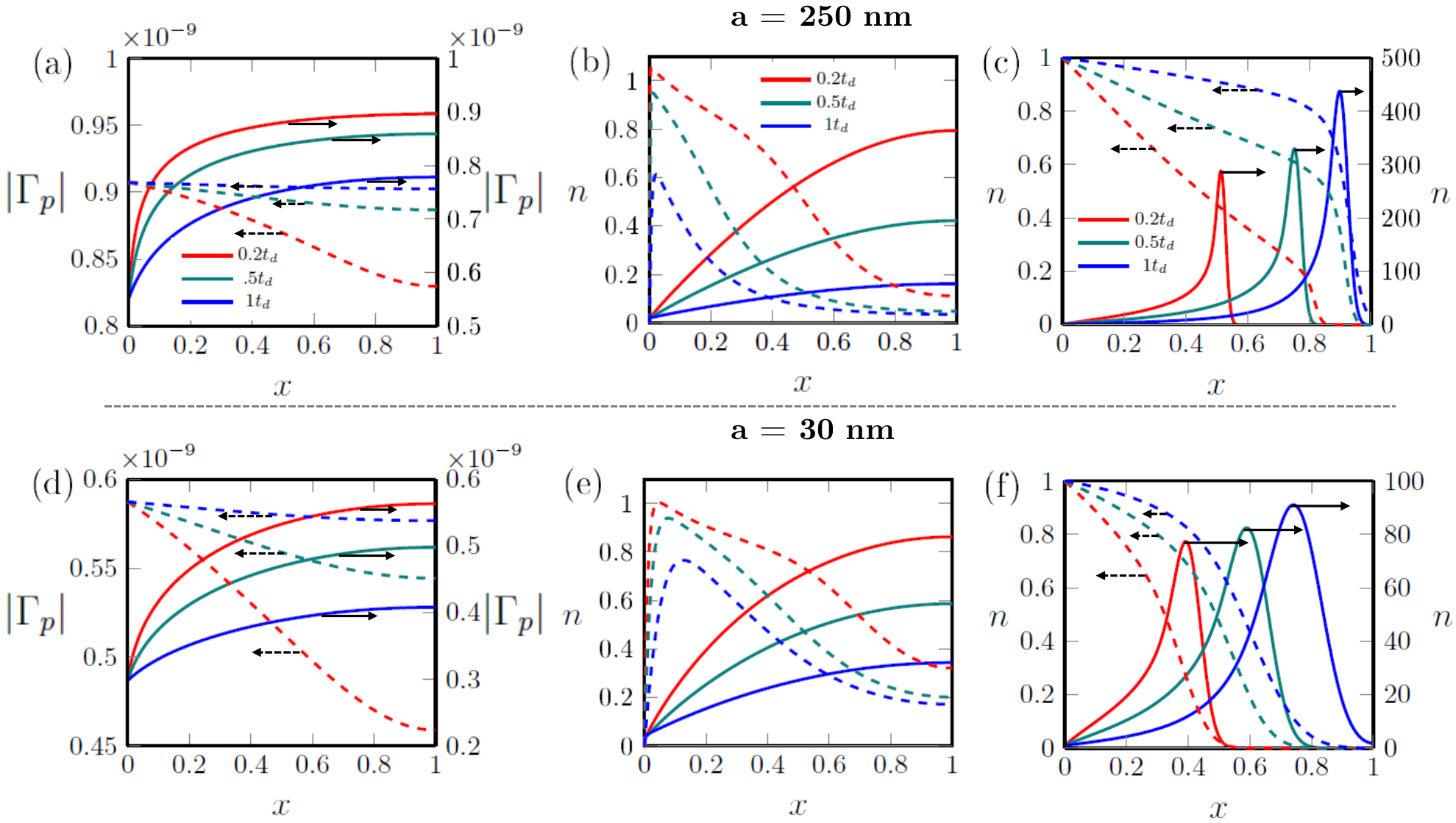}}
    \caption{(a) and (d) Particle mobility ($\Gamma_P$, m$^2$/s) variation along the length with Debye layer effect at $t^*=0.2t_d, 0.5t_d$ and $1t_d$ for $a=250$ nm and $30$ nm, respectively. (b,c)  Particle withdrawal and (e,f) injection patterns for solute-out (solid line) and solute-in mode (dashed line) at $a=250$ nm and $30$ nm, respectively. Parameters: $D_s=1\times 10^{-9}$ m$^2$/s, $D_p=1\times 10^{-12}$ m$^2$/s ($a=250$ nm) and $8.084\times 10^{-12}$ m$^2$/s ($a=30$ nm), $\beta=0.01$ (S-O) and $100$ (S-I).}
    \label{fig:E2}
    \end{figure*}

\begin{figure*}[t]
    \centering
    {\includegraphics[width=0.8\linewidth]{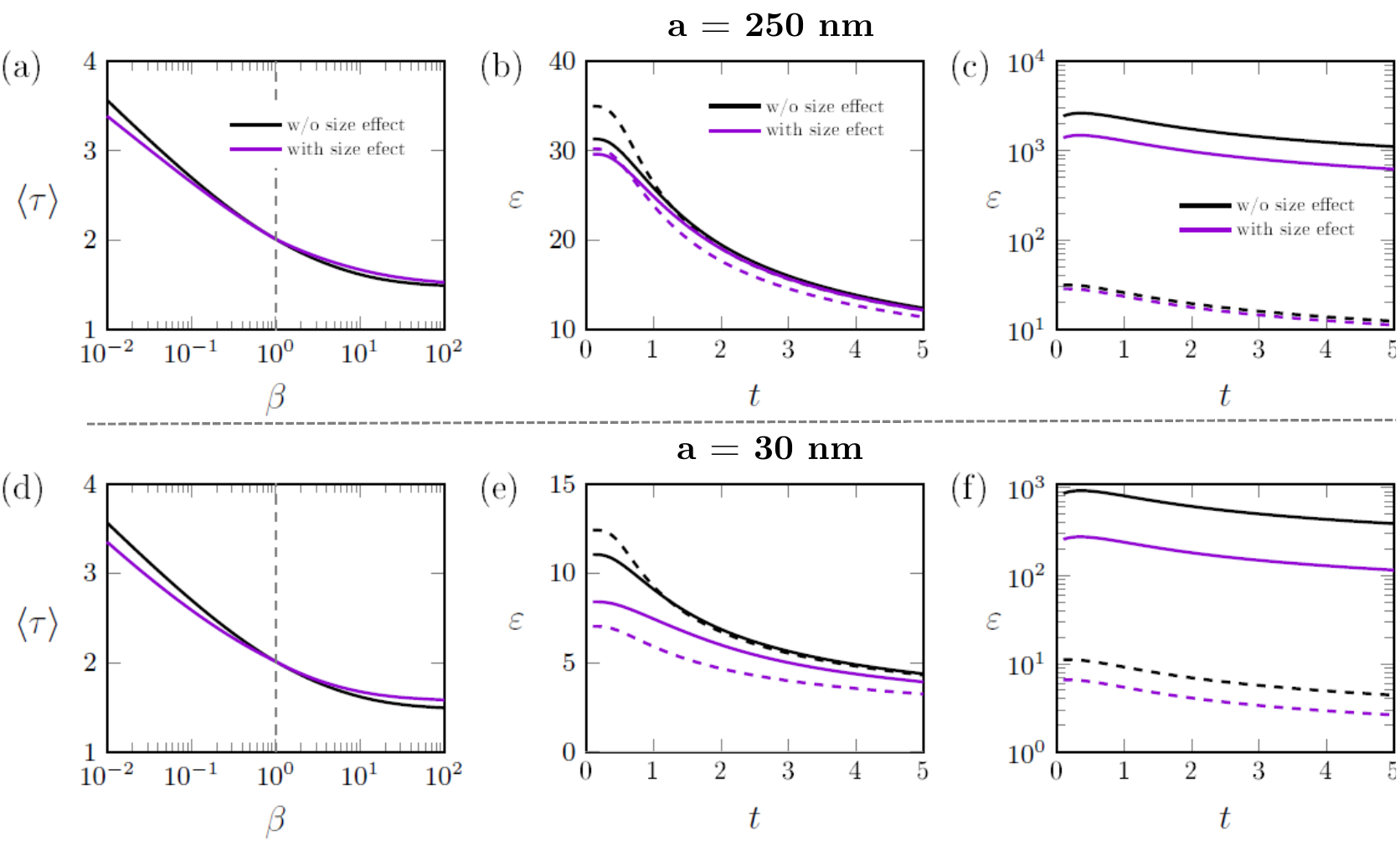}}
    \caption{(a) and (d) Persistence time variation with applied concentration
difference ratio $\beta$ with and without Debye later effect for $a=250$ nm and $30$ nm, respectively. Effectiveness of (b,c) withdrawal and (e,f) injection cases for $a=250$ nm and $30$ nm, respectively. Parameters: $D_s=1\times 10^{-9}$ m$^2$/s, $D_p=1\times 10^{-12}$ m$^2$/s ($a=250$ nm) and $8.084\times 10^{-12}$ m$^2$/s ($a=30$ nm), $\beta=0.01$ (S-O) and $100$ (S-I).}
    \label{fig:E3}
    \end{figure*}

\section*{Appendix}

\rev{
When colloids are of sub-micron \& nanometer size, the typical ($\sim 1-50$ nm) Debye layer (DL) can no longer be considered as infinitesimally thin compared to colloid radius.
Here we utilize the mobility corrections from \cite{keh2000diffusiophoretic,ganguly2023diffusiophoresis} to account for the finite DL effects to predict the spatiotemporal patterns and performance. The phoretic mobility is modified as follows:
\begin{equation}
   \Gamma_p = \frac{\varepsilon}{\mu} \left[ \frac{k_b T}{e} \beta_s \zeta \, \Theta_1\left(\frac{a}{\lambda}\right) + \frac{\zeta^2}{8} \, \Theta_2\left(\frac{a}{\lambda}\right) \right]
\end{equation}
The modification functions 
\begin{align}
    \Theta_1(\xi) &= 1 - \frac{1}{3} \left(1 + 0.072\, \xi^{1.13} \right)^{-1}, \\
    \Theta_2(\xi) &= 1 - \left(1 + 0.085\, \xi + 0.02\, \xi^{0.1} \right)^{-1}
\end{align}
depend on the particle radius ($a$) and the DL thickness ($\lambda$). Expanding $\lambda$ using the Gouy–Chapman theory
($\lambda = \sqrt{\frac{k_B T \varepsilon}{2 N_A \mathcal{C} e^2}}$), where $\mathcal{C}$ is the molar concentration (i.e., $c^* \times 10^3$ mol/m\textsuperscript{3}), reveals that $\lambda$, and hence the modification functions, are functions of the local solute concentration.
As concentration increases, the Debye layer becomes thinner and phoretic mobility correspondingly increases.
The figures below present simulations in S-O and S-I modes, wherein solute concentrations span a broad range ($0.1-10$ mM). 
Specifically, in the S-O mode, the ambient concentration is set to 0.1 mM while the initial pore concentration is 10 mM; the reverse configuration is used for the S-I mode.
For a 250 nm colloid, Fig. A1 (a) shows the variation of mobility across the range of concentrations considered. For a smaller particle, the gap between the (ideal) constant mobility and variable mobility is larger (Fig. A1(d)).
In Figs. A1 (b,c), we find that the constant mobility model overestimates the colloidal withdrawal and injection. Figs. A1 (e,f) show how the DL effects further dominate for smaller sized colloids.
Fig. A2 compares the two modes for thick DL model. Fig. A2 (a) shows the spatiotemporal variation of mobility in the pore; we note that the mobility rises with pore depth for S-O mode as the concentration increases with $x$, whereas a decrease is noted for S-I mode.
Comparing the y-axes, we also note that, beyond $t^*\gtrsim t_d$, the mobility for the S-I mode does not decay significantly in space, as mobility varies weakly with increasing concentration. This can also be interpreted from Fig. A1(a), where mobility variations in the vicinity of $c^* = 0.1$ mM are more pronounced, whereas they are much weaker in the vicinity of $c^* = 10$ mM.
Figs. A2(b,c) show that the spatiotemporal patterns of withdrawal and injection follow trends qualitatively similar  to those observed in the constant mobility model. Figs. A2(d–f) present analogous results, with the thick DL effect further damping the injection and withdrawal patterns.
}

\rev{In the context of performance, Fig. A3(a,d) shows that the persistence time exhibits a trend similar to that of the constant mobility model. It can also be observed that the persistence in the S-O mode is reduced,  this is because of substantial decrease in mobility, particularly near the pore mouth, as seen in Fig. A2(a,d). Additionally, the gradual decline in concentration over time further lowers the mobility. In contrast, for the S-I mode, persistence is slightly enhanced, as the rising concentration over time leads to an increase in mobility, consistent with Fig. A2(a,d).
 Fig. A3 (b,e), (c,f) show that, relative to the constant mobility model, the thick DL effects dampen the effectiveness trends of withdrawal and injection, respectively.
It is interesting to note a finer aspect of withdrawal effectiveness for extremely large solute gradients: in the initial stages, the S-I mode can exhibit slightly enhanced effectiveness compared to the S-O mode, likely due to the nature of spatial pattern at early times. For the S-I mode, the spatial non-monotonicity of $\IB{u}_p$ renders near-pore-mouth accumulation, which readily convects and diffuses out due to large curvature. 
For thick DL mobility model, this brief dominance of the source–injection (S–I) mode at early times is attenuated due to reduced mobility.}

\section*{Associated Content}
\textbf{Supporting Information:}  Validation of results and further information on constant charge model, diffusio-osmotic flow, and quantification of distribution spread during colloidal injection.

\section*{Acknowledgements}
K.T thanks Rohit Bhattacharjee from MathWorks for aid in parallel computation of results. J.D thanks the IIT Kanpur Institute Postdoctoral Fellowship. A.C acknowledges the funding from IIT Kanpur Initiation Grant. 
The authors thank Ankur Gupta from University of Colorado Boulder for his suggestions. 
The authors also thank the referees for their comments on improving the manuscript.

\small
	\bibliography{ref_modified}
\end{document}